\documentclass[conference]{IEEEtran}
\IEEEoverridecommandlockouts
\usepackage{cite}
\usepackage{amsmath,amssymb,amsfonts}
\usepackage{algorithmic}
\usepackage{graphicx}
\usepackage{textcomp}
\usepackage{xcolor}
\def\BibTeX{{\rm B\kern-.05em{\sc i\kern-.025em b}\kern-.08em
    \kern-.1667em\lower.7ex\hbox{E}\kern-.125emX}}
\usepackage{balance}
\usepackage{cite}
\usepackage{amsmath,amssymb,amsfonts}
\usepackage{algorithmic}
\usepackage{graphicx}
\usepackage{textcomp}
\usepackage[ampersand]{easylist}
\usepackage{amssymb}
\usepackage{multicol}
\usepackage{multirow}
\usepackage{tabularx}
\usepackage{rotating}
\usepackage[roman]{parnotes}
\usepackage[many]{tcolorbox}
\usepackage{soul}
\usepackage{xcolor}
\usepackage{comment}
\usepackage{booktabs}
\usepackage{colortbl}
\usepackage{verbatim}
\usepackage{hyperref}
\usepackage{array}
\usepackage[inline]{enumitem}
\usepackage[export]{adjustbox}
\usepackage{subfig}

\newcolumntype{L}[1]{>{\raggedright\let\newline\\\arraybackslash\hspace{0pt}}m{#1}}
\newcolumntype{C}[1]{>{\centering\let\newline\\\arraybackslash\hspace{0pt}}m{#1}}
\newcolumntype{R}[1]{>{\raggedleft\let\newline\\\arraybackslash\hspace{0pt}}m{#1}}

\newif\ifdraft
\draftfalse 

\newcommand{\boldification}[1]{\ifdraft\indent ** \textbf{#1} **\\\indent\else\relax\fi}

\begin{document}
\title{A Decade of Information Architecture in HCI: A Systematic Literature Review}

\author{
\IEEEauthorblockN{Mariam Guizani}
\IEEEauthorblockA{\textit{Department of EECS, Oregon State University} \\
guizanim@oregonstate.edu}
}

\maketitle
\begin{abstract}
Information Architecture (IA) is a blueprint for the information system in websites or other information-rich environments. It corresponds to how we organize, label and structure information. The importance of Information Architecture and its influence on a system's usability is vastly discussed in literature. 

Because of the inherent connection between Information Architecture concepts and the Human Computer Interaction (HCI) field, we decided to investigate how previous research has used Information Architecture in the context of Human Computer Interaction (IAinHCI). 

In order to do that, we followed a two phase process. First, we conducted a  Systematic Literature Review (SLR). We queried both the ACM and IEEE databases. We filtered and assessed 311 papers that spanned a decade of research on Information Architecture. We found 25 papers that utilized Information Architecture in the context of Human Computer Interaction. Then, we followed a Background Reference Search process using the SLR resulting papers as a starting set. We assessed the eligibility of the reference list of all 25 papers and found eight additional papers that were relevant to our research question. 

Results of our review show that, IAinHCI papers fall under seven main categories, from IoT to the semantic web and ubiquitous technology. The website category, however, was both the most consistent over the years and the most prevalent category accounting for 67\% of the papers. Our findings suggest that IA has not yet uncovered its full potential and there is still room for research to leverage and expend the IA knowledge base promising a prosperous future for Information Architecture.
\end{abstract}

\begin{IEEEkeywords}
Systematic Literature Review, Information Architecture, Human Computer Interaction
\end{IEEEkeywords}

\section{Introduction}
\label{sec:intro}
In today's information rich environment, the way information is structured and organized plays an important role in task completion  \cite{garrett1900elements}\cite{morville2006information}. Information Architecture is the art and science of organizing, labeling and structuring information \cite{morville2006information} to help users understand their surroundings and assist them in finding information. As Malonery et al.\cite{maloney2004beyond} put it, `` Web-site design adds  presentation and graphical elements to IA to create the user  experience.'' Thus, IA sets the foundation for an effective communication and interaction. 

Because of the Information Architecture's influence on the user experience\cite{nielsen2006prioritizing}, we decided to investigate previous research on Information Architecture in the context of Human Computer Interaction. Our main goal is to understand how IA concepts have been leveraged in HCI to study, evaluate and improve the user experience. 

In order to achieve our objective, we investigated a decade of Information Architecture research in HCI by first conducting a Systematic Literature Review (SLR)\cite{kitchenham2004procedures} querying both the ACM and IEEE databases and then complementing the SLR results with a Backward Reference Search \cite{skoglund2009reference}. 

Our results give an overview of the different topics and research categories covered by IAinHCI with the purpose of (1) understanding the current state of the art and (2) uncovering research opportunities that can leverage and build upon the Information Architecture knowledge base.

\section{Background}
The word information architecture was first coined in the mid-1970's by Richard Saul Wurman\textemdash an architect and a web designer\textemdash as a way of ``making the complex clear''. Information Architecture is an interdisciplinary field with definitions that vary from ones providing a clear delimitation that view IA as a set of organization, labeling and navigation systems \cite{morville2006information} to broader definitions were IA is described as the design of the information space as a whole \cite{dillon2005information}. Even though there is an absence of a formal definition, the fact that IA has an impact on a system's usability appear to be a central thought \cite{morville2006information} \cite{nielsen2006prioritizing} \cite{garrett1900elements} \cite{maloney2004beyond}.

The seminal work of Rosenfeld and Morville \cite{morville2006information}, often referred to as the ``bible'' of IA, divides IA into four different system: The organization system, the navigation system, the labeling system and the search system. 
\paragraph{The Organization System} is composed of both an organization scheme and an organization structure \cite{morville2006information}. 
The organization scheme describes the rationale behind the grouping of content. An exact organization scheme (e.g, chronological , alphabetical, geographical) is used when the end user know exactly what they are looking for. In most other cases, a subjective scheme (e.g, topic, task, audience, metaphor) is employed. For instance, a task based organization scheme would have to accommodate the high priority tasks that the target users would perform. The organization structure, as the name implies, is what defines the relationship between content groups. Hierarchy and hypertext are two examples of an organization structure.

\paragraph{The Navigation System}
As  Rosenfeld and Morville \cite{morville2006information} describe it  ``Structure and organization are about building rooms. Navigation design is about adding doors and windows.'' The navigation system is what enables the user to traverse the information structure. The navigation system is divided into two main categories: the embedded navigation system and the supplemental navigation system. While the embedded navigation system (e.g, global menus, local menus, contextual links) are inherent to the hierarchical structure of an information system, the supplemental navigation system (e.g, site maps, guides, site indexes) is extrinsic to a system's basic structure.

\paragraph{The Labeling System} 
The labeling system is closely tied to the organization system. Meaning the labels have to match the site's organization. Labels are relevant to a variety of Information Architecture elements. For example labels can be applied to contextual links, headings and index terms. Often times users perceive labels as cues that guide their navigation, thus it is important to avoid jargon and favor user-centric labels \cite{morville2006information}.

\paragraph{The Search System} 
The search system could be thought of as a supplemental system. It allows the user to retrieve information using a particular term or phrase \cite{morville2006information}. Not all information systems need a search feature. Thus, determining whether a specific site can benefit from a search system is a critical preliminary step to unsure that the end user is not overwhelmed when there is a considerable amount of information. For example, systems with highly dynamic content can benefit from having a search feature. First, a search system would help with site maintenance by automatically indexing the variety of content that could be changing daily and second, it would provide a feature that, in such a case, would be highly expected by the end user. 

In this paper we will adopt the aforementioned definition and set of systems as described by Morville and Rosenfeld \cite{morville2006information}.

\section{Methodology}
\label{sec:method}
In this paper we first follows a Systematic Literature Review process as described by Kitchenham \cite{kitchenham2004procedures}. We then complement our SLR findings \cite{keele2007guidelines} by performing a Backward Reference Search  \cite{skoglund2009reference} \cite{wohlin2014guidelines} using the final SLR included papers as a starting set.

We set the rationale for the study as mean to identify available research that leverages Information Architecture (IA) in an HCI context (IAinHCI). 
We frame our SLR around a central research question:
\begin{itemize}
    \item How was Information Architecture leveraged in the context of Human Computer Interaction?  
\end{itemize}


In order to investigate our central research question we follow a total of a nine steps process (Main SLR: 5 steps and Backward Reference Search: 4 steps)

\subsection{Main SLR}
\begin{figure}[bthp]
\graphicspath{ {figures} }
         \includegraphics[width=.45\textwidth]{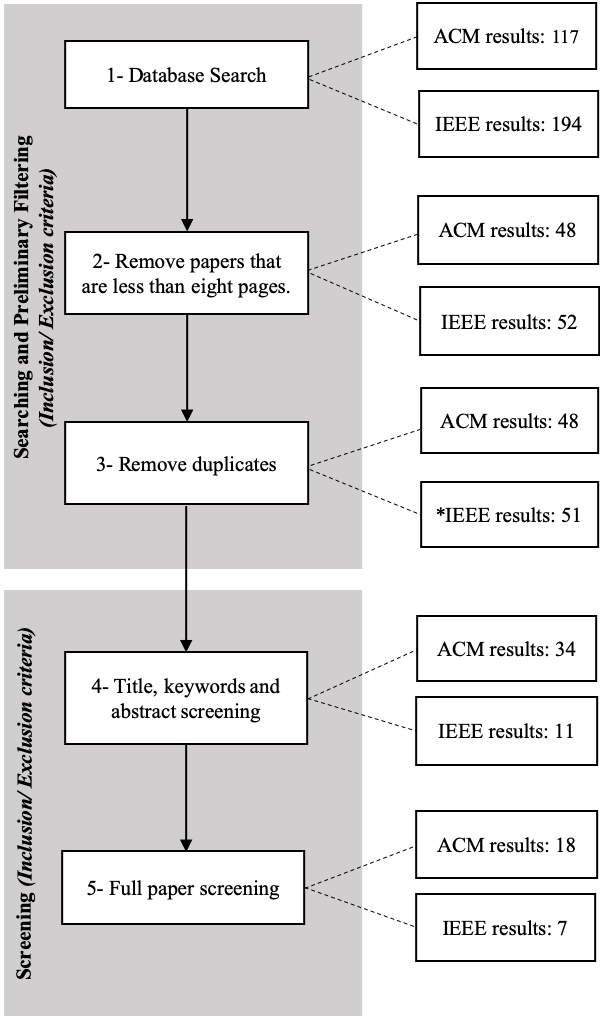}
\caption{The main SLR review process. The * result is the one from which the duplicate was removed.}
\label{fig:SLRmain}
\end{figure}

The main SLR search [Figure \ref{fig:SLRmain}] consisted of separately querying both the ACM and IEEE databases using our search string and then, screening for the papers that are relevant to our research question.

\begin{figure}[bthp]
\graphicspath{ {figures} }
         \includegraphics[width=.45\textwidth]{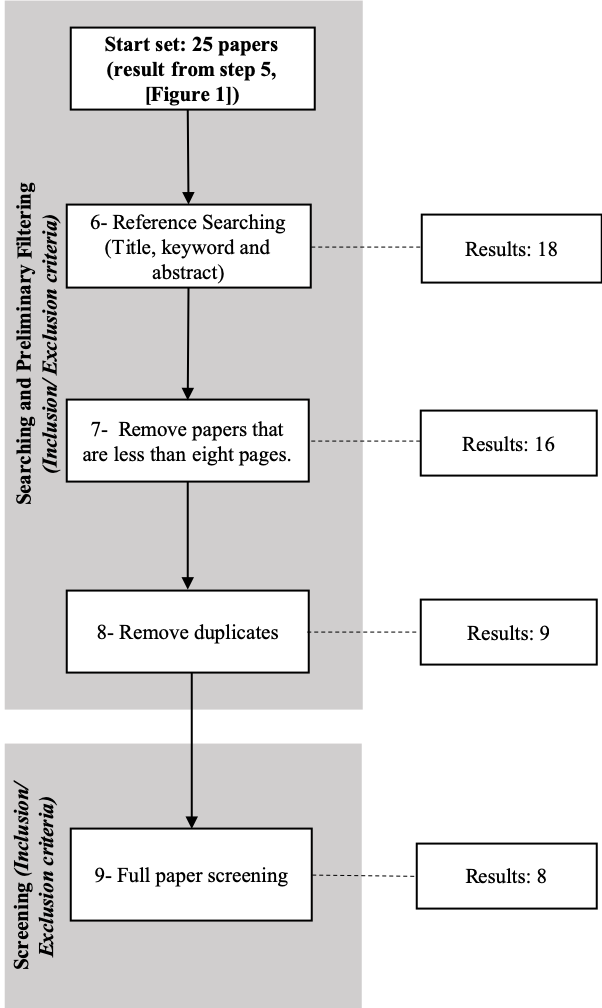}
\caption{The Backward Reference Search process.}
\label{fig:Refsearch}
\end{figure}

\subsubsection {Step 1, 2 \& 3: Searching and Preliminary Filtering}
We defined our search string as follow:
\begin{itemize}
    \item``Document Title'':``information architecture'' OR ``Abstract'':``information architecture'' OR ``Author Keywords'':``information architecture'' 
\end{itemize}
We filtered the ACM and IEEE databases output to capture the last decade of IAinHCI research. 
\begin{itemize}
    \item ``publication Year'': 2009, 2019
\end{itemize}

We designed our search string in a way that ensures the relevance of the papers captured (The term ``information architecture'' has to appear at least once in the title, abstract or keywords). In order to avoid missing on important papers, we decided to manually filter out those where Information Architecture was not used in an HCI context. 

In order to make sure we are only capturing non redundant fully developed paper ideas, we follow an inclusion/ exclusion criteria\textemdash  removing duplicate and papers that were less than eight pages long. This threshold has been defined by researching different top tier venues specifications on what defines as a full paper and by running a pilot study. 
Only one duplicate was found between the two databases which we removed from the IEEE search output [Step 3, Figure \ref{fig:SLRmain}].

\subsubsection{Step 4 \& 5: Screening}
We screened the titles, abstracts and keywords (Step 4, [Figure \ref{fig:SLRmain}]) of a total of 99 papers (48 ACM, 51 IEEE). We followed a set of inclusion/ exclusion criteria, removing: 
\begin{itemize}
    \item Documents that were not papers: extended abstracts, magazines, books and book chapters.
    \item Papers written in a language other than English.
    \item Papers that are not relevant to IAinHCI: papers that are not related to using the Information Architecture lens in order to design a product, evaluate its usability or understand its underlying structure. For example, a paper that introduces an automated tool for card sorting or one that describes the roles of information architects are filtered out. 
\end{itemize}

In the event where a paper's abstract was unclear or presented incomplete information, the paper was kept and eliminated during the full text read (Step 5, [Figure \ref{fig:SLRmain}]) if justified. 

The full text read (Step 5, [Figure \ref{fig:SLRmain}]) was conducted by first reading the abstract and conclusion of the paper in order to get an idea on what the paper is about and what the main findings are, followed by a screening of the method to understand how the study was conducted and then a reading of the results section to get a deeper understanding of the contribution of the paper. 

\subsection{Backward Reference Search}
As stated in \cite{skoglund2009reference} and \cite{keele2007guidelines}, an SLR should be complemented by a reference search. In order to do so, we performed a backward reference search [Figure \ref{fig:Refsearch}] and followed the same inclusion/ exclusion criteria mentioned in the SLR method. We selected the final papers resulting from the SLR process as our starting set. We then scanned all the reference titles. Papers whose title proved relevant to our research question were further scanned using the abstract and keywords [Figure \ref{fig:Refsearch}]. Papers that had at least one occurrence of the phrase ``information architecture'' in the title, keyword or abstract were kept for further screening. 
After removing the papers that were less than eight pages long as well as duplicate papers, we performed a full text read of the remaining articles.

\subsection{Data Collection}
Results of both the main SLR and the Backward Reference Search were stored in an excel sheet: The metadata stored consisted of:
\begin{itemize}
    \item Publication title.
    \item Author list.  
    \item Publication year.  
    \item Author's keywords. 
    \item Number of pages. 
    \item Paper link.
\end{itemize}

An additional decision column was added were the decision of keeping or removing the paper was mentioned along with a justification. 

Two extra columns were added during the full text read: a category column and a summary column. 

\section{Results}
\label{sec:results}
\boldification{number of papers screened, assessed for eligibility }
\begin{figure*}[ht]
\graphicspath{ {figures} }
         \includegraphics[width=.97\textwidth]{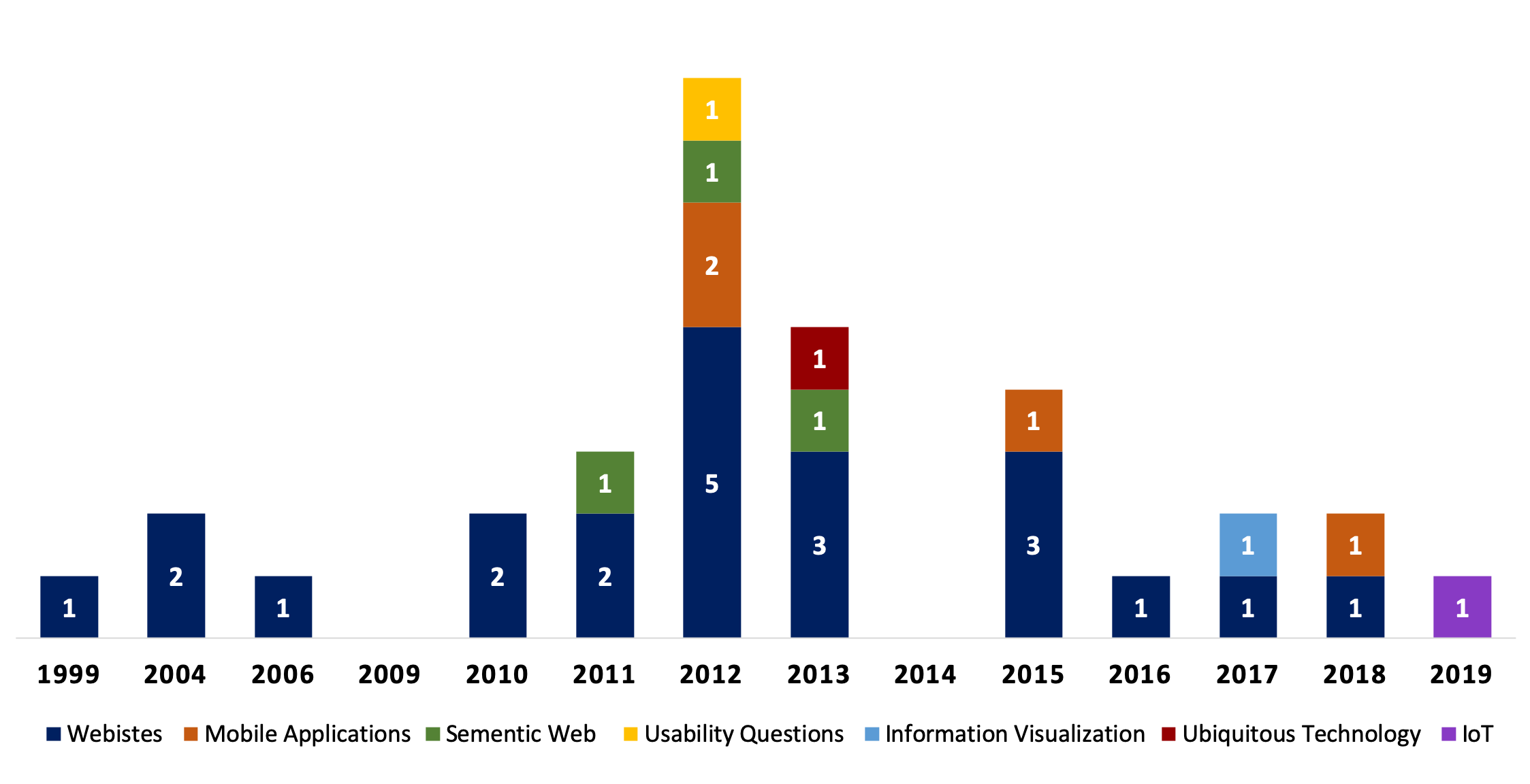}
\caption{The categories of the final set of papers (results of the SLR and the Backward Reference Search) per year of publication (years before 2009 are omitted if they are not associated with a paper).}
\label{fig:categperyear}
\end{figure*}
In this section, we will first, provide the number of papers assessed for eligibility during the SLR and the Backward Reference Search steps. Then, we will answer our central research question :  How was Information Architecture leveraged in the context of Human Computer Interaction? 
In order to do so, we break down our central research question into three questions (1) What are the main categories that have been researched and what is their distribution over time? (2) How was IA used in the most prevalent category? and (3) How was IA used in the not so prevalent categories? 

\subsection{SLR and Backward Reference Search Results}
The query from the ACM and IEEE databases returned 311 papers (117 ACM and 194 IEEE) that spanned a decade of research. First, a preliminary filtering (Steps 2 \& 3, [Figure \ref{fig:SLRmain}]) resulted in 48 ACM papers and 51 IEEE papers. Only one duplicate was found between the two databases which we removed from the IEEE search output. Out of the 99 candidate articles whose title, keyword and abstract were screened (Step 4, [Figure \ref{fig:SLRmain}]), 45 remained (34 ACM papers and 11 IEEE papers). The full text read (Step 5, [Figure \ref{fig:SLRmain}]) resulted in our final SLR set of 25 papers (18 ACM papers and 7 IEEE papers). 

The Backward Reference Search resulted in 18 candidate papers after the title, abstract and keyword filtering (Step 6, [Figure \ref{fig:Refsearch}]). The remaining nine papers from step 7 \& 8 [Figure \ref{fig:Refsearch}] were assessed with a full text read. The aforementioned last step (Step 9, [Figure \ref{fig:Refsearch}]) resulted in the removal of an additional paper.  

The SLR main search and the Backward Reference Search resulted in a total of 33 papers related to our research question. 
\begin{figure}[htb!]
\graphicspath{ {figures} }
         \includegraphics[width=.45\textwidth]{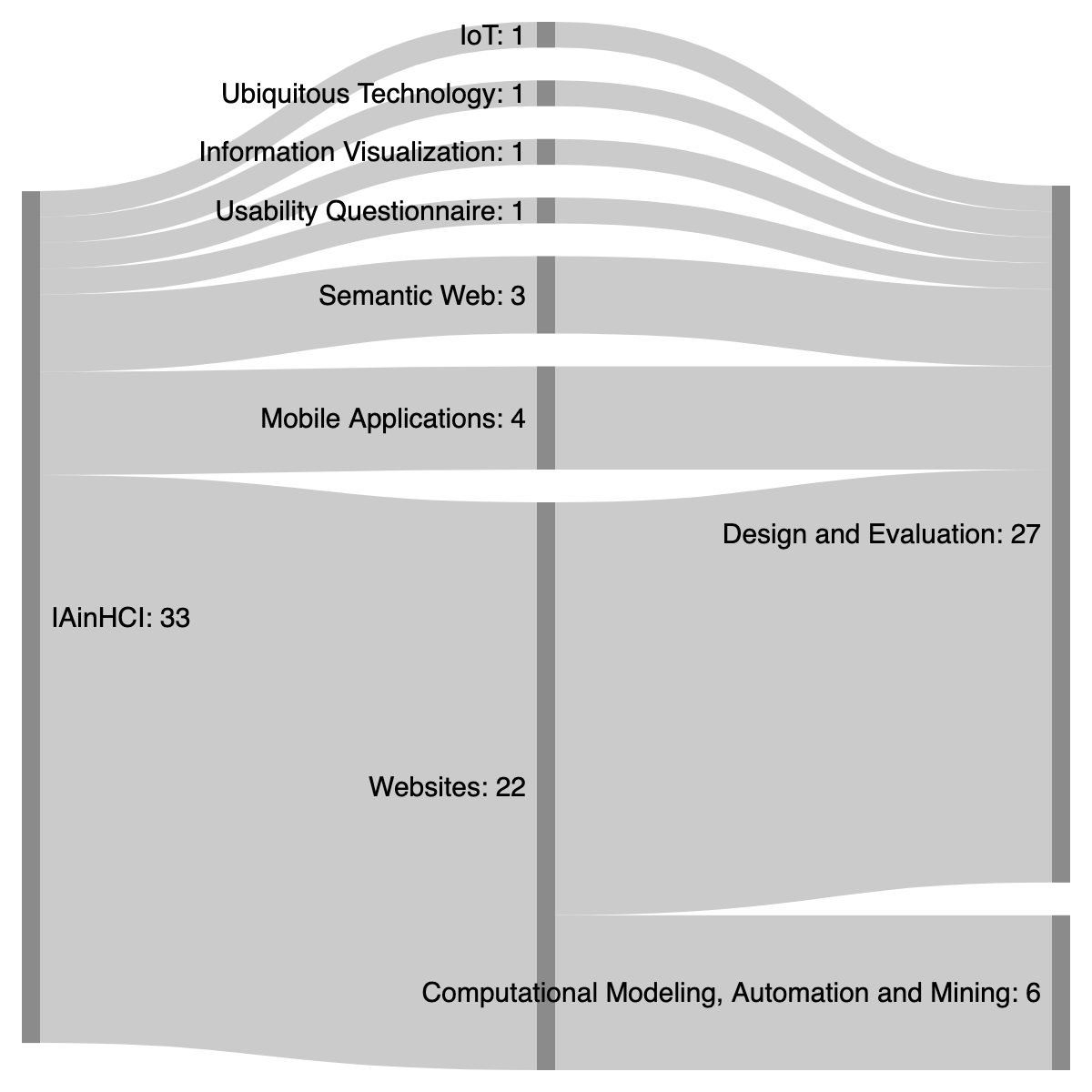}
\caption{The main categories of research literature on Information Architecture in the context of Human Computer Interaction (IAinHCI).}
\label{fig:CompleteflowCateg}
\end{figure}

\subsection{The Different Categories and Their Distribution Over The Years}
The resulting papers fall under seven main categories [Figure \ref{fig:CompleteflowCateg}].
Our results show that Information Architecture is mostly used to study websites. In fact, 67\% of our final set of papers fall under the website category [Figure \ref{fig:CompleteflowCateg}] making it, by far, the most prevalent category. This is consistent with the seminal IA literature that focuses on IA for the world wide web \cite{morville2006information}. The second most common category is mobile applications accounting for 12\% [Figure \ref{fig:CompleteflowCateg}] of our final set of papers.

The majority of the reviewed papers (82\%) researched IA driven design and evaluation [Figure \ref{fig:CompleteflowCateg}]. The remaining 18\% of the papers used IA for computational modeling, automation and mining.

Figure \ref{fig:categperyear} display the category distribution of our final set of papers per year of publication. All categories combined, the 2011-2013 year range accounts for 52\% of all the IAinHCI papers in our set. Additionally, all the semantic web related papers cluster in that range and 45\% of the papers within the website category have been published between 2011 and 2013. IAinHCI for website appears to be a somewhat constant area of research except in 2009 and 2014 were we notice a gap. In contrast, both IAinHCI in the Information visualization category (one paper published in 2017) and IAinHCI in the  IoT category (one paper published in 2019) represent more recent areas of research.  

The year 2012 is associated with a peak in IAinHCI publications (27\% of the papers). More specifically, 2012 is also associated with the peak publication in the website category (23\% of all website related papers have been published in 2012). 
\subsection{The Prevalent Category: Websites}
\begin{figure}[htb!]
\graphicspath{ {figures} }
         \includegraphics[width=.45\textwidth]{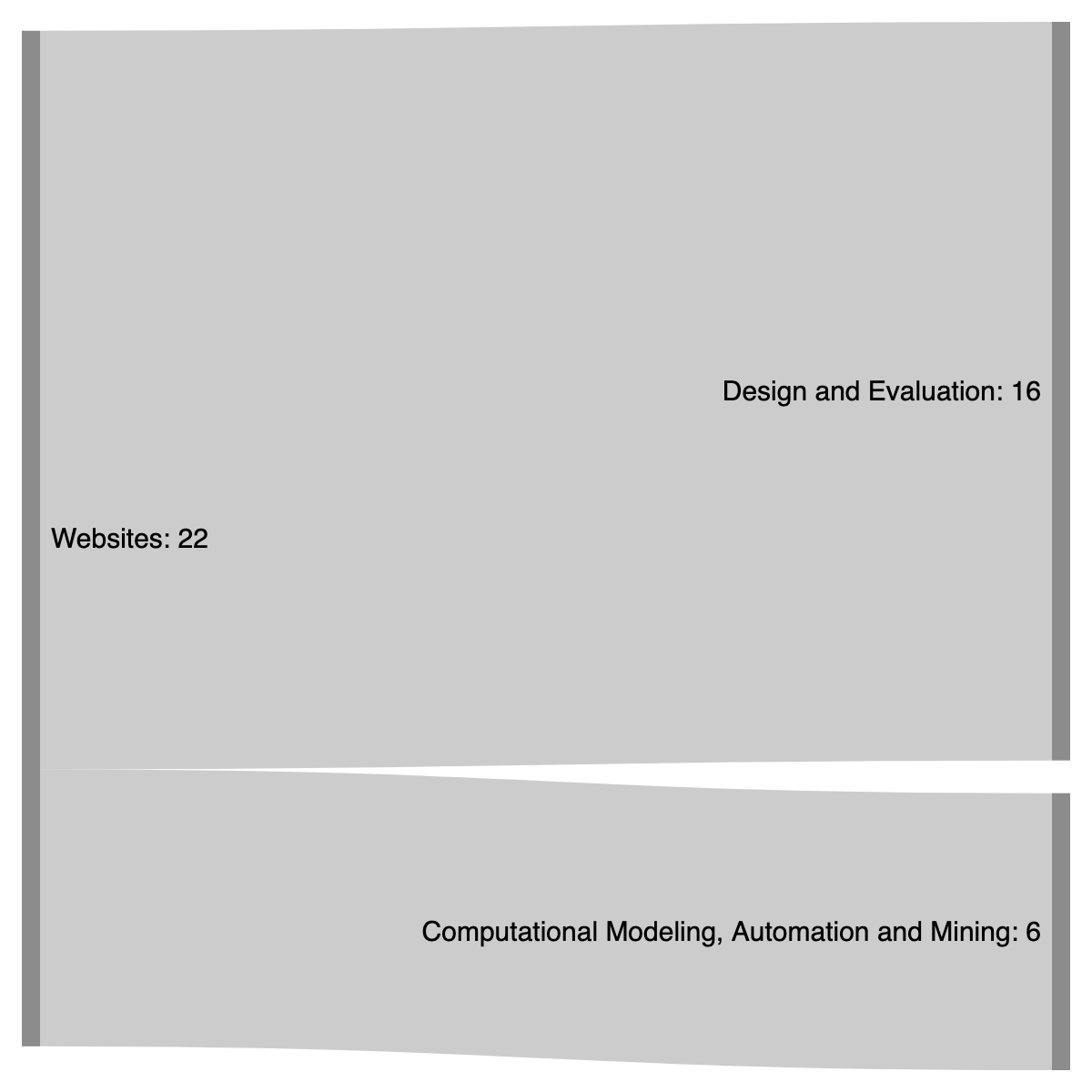}
\caption{The websites subcategories.}
\label{fig:WebsiteGeneralFlow}
\end{figure}

 The usage of IA for websites falls under two main subcategories. (1) The design and evaluation of websites and (2) the computational modeling, automation and mining of websites [Figure \ref{fig:WebsiteGeneralFlow}]. 73\% of the papers categorized as website related, use IA in their design and evaluation process [Figure \ref{fig:WebsiteGeneralFlow}].
\subsubsection{Design and Evaluation}
\begin{figure}[h]
\graphicspath{ {figures} }
         \includegraphics[width=.45\textwidth]{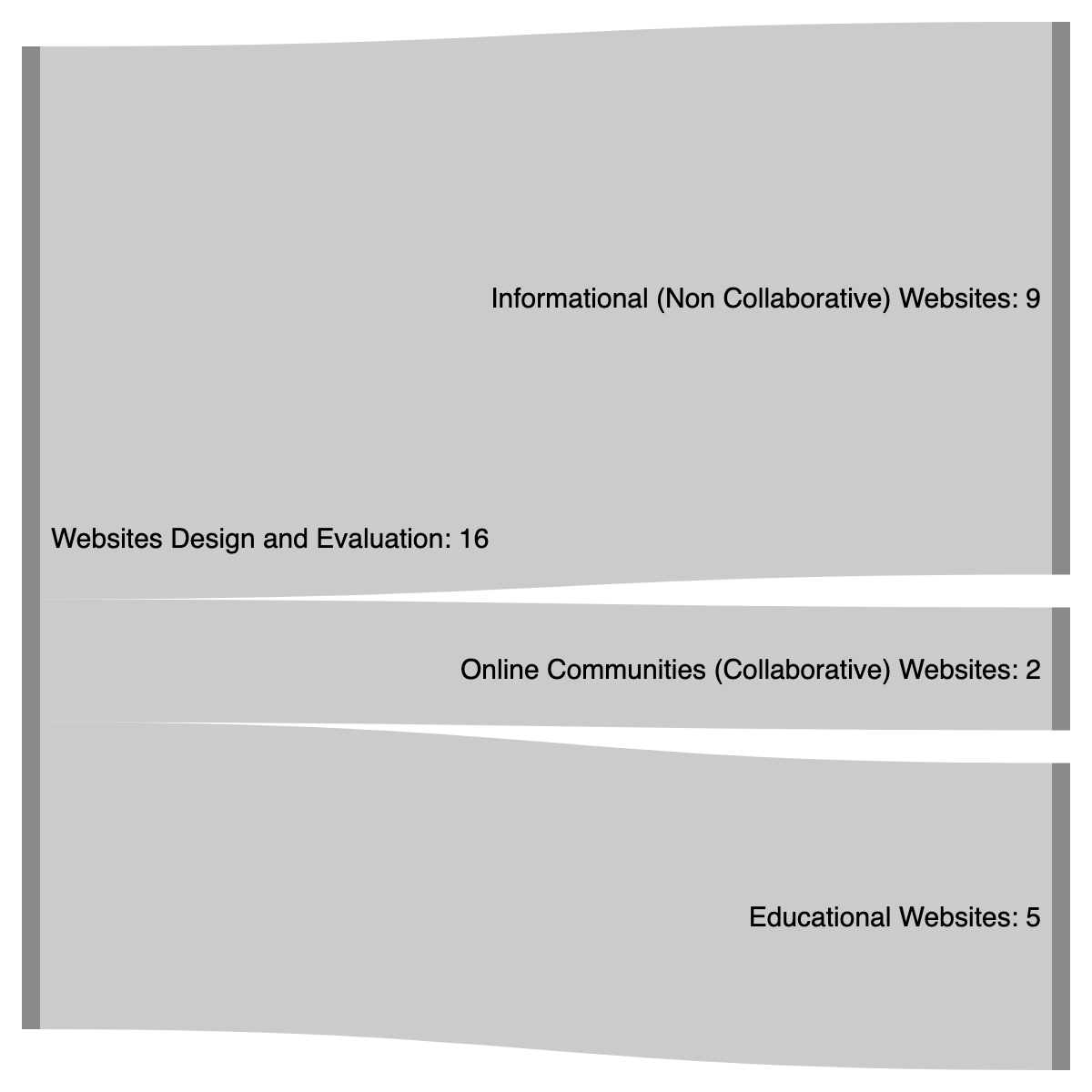}
\caption{The finer grained groups of the websites design and evaluation subcategory.}
\label{fig:D&Ewebsite}
\end{figure}
Papers in the design and evaluation subcategory fall under three different groups [Figure \ref{fig:D&Ewebsite}]

\paragraph{Informational (Non Collaborative) Websites}
Multiple research investigated IA in informational and non collaboration websites \cite{aranyi2012using}\cite{petrie2012users}
\cite{resnick2004effects} \cite{wang2012critical}.

Both study \cite{aranyi2012using} and \cite{petrie2012users} showed that Information Architecture is a concern that users voiced when interacting with the evaluated websites. In fact, Aranyi et al. \cite{aranyi2012using} evaluation of a news website showed that IA is the second most prevalent category after content and Petri and Power \cite{petrie2012users} findings showed that Information Architecture accounted for 9\% of the issues reported by both users and experts. 
In contrast, Resnick and Sanchez \cite{resnick2004effects} focused on a subset of the Information Architecture schemes (The organizational and labeling scheme). They conducted a user study with 60 participants to evaluate the effect of the organization and labeling of information on the performance and user satisfaction in the context of an e-commerce website. Results of the study showed that user-centric labels significantly benefited the performance and user satisfaction. The user-centric organization, however, resulted in an improved performance only when combined with low quality labels. 

The informational non collaborative websites are widely used by diverse users on a daily basis. As stated by the aforementioned research, the Information Architecture plays a role in this daily interaction. Swierenga et al. \cite{swierenga2011website} study  goes beyond the visual IA to compare the experience of sighted, visually impaired and user with low vision while using an outdoor recreational website. The empirical findings displayed a drastic difference in task duration between sighted and visually impaired individuals (3-4 times longer). Many of the barriers encountered by both visually impaired and low vision individuals had to do with the organization and the labeling system. 
Authors of \cite{swierenga2011website} bring the study to a close by providing general and IA driven guideline for both sighted and visually impaired users. In Alignment with Swierenga et al. study \cite{swierenga2011website}, a multitude of research\cite{bolchini2006designing} \cite{rohani2013mobile} \cite{yang2012aural} \cite{vigo2013challenging} \cite{rohani2012back} was conducted on the design and evaluation of aural Information Architecture (aural IA). For instance, Bolchini et al. \cite{bolchini2006designing} gave an overview of potential design issues when designing an aural Information Architecture and provided some guidelines inspired by their experience in the development of applications for visually impaired users. Ghahari et al. \cite{rohani2013mobile} introduced and evaluated an aural Information Architecture framework for the web: aural flow. Built on top of an existent news website, the goal of this framework is to provide a navigation system with minimal screen interaction that support both breadth first and depth first navigation. 

\begin{figure}[bthp]
\centering
\graphicspath{ {figures} }
         \includegraphics[width=.48\textwidth]{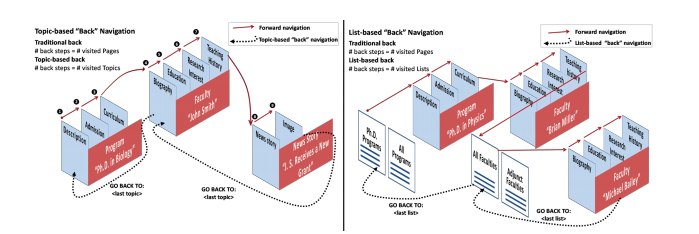}
\caption{Topic and list based back navigation as designed by \cite{rohani2012back}. Topic back navigation (left) where users can navigate to previously visited topics. List based back navigation (right) enables users to go to previously visited lists.}
\label{fig:auralIAtopicVSlist}
\end{figure}

Ghahari et al. study \cite{rohani2012back} introduced two aural back navigation strategies\textemdash a topic based and a list based strategy [Figure \ref{fig:auralIAtopicVSlist}] and evaluated them against the traditional one page at a time back navigation. Their results suggest that both the topic and list based back navigation strategy enhanced the navigation effectiveness and efficiency. 

\begin{figure}[bthp]
\centering
\graphicspath{ {figures} }
         \includegraphics[width=.35\textwidth]{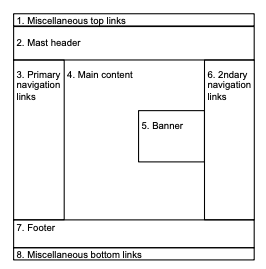}
\caption{The stereotypical layout of a web page as presented in \cite{vigo2013challenging}.}
\label{fig:steriolayout}
\end{figure}

\begin{figure*}[bthp]
\centering
\graphicspath{ {figures} }
         \includegraphics[width=.95\textwidth]{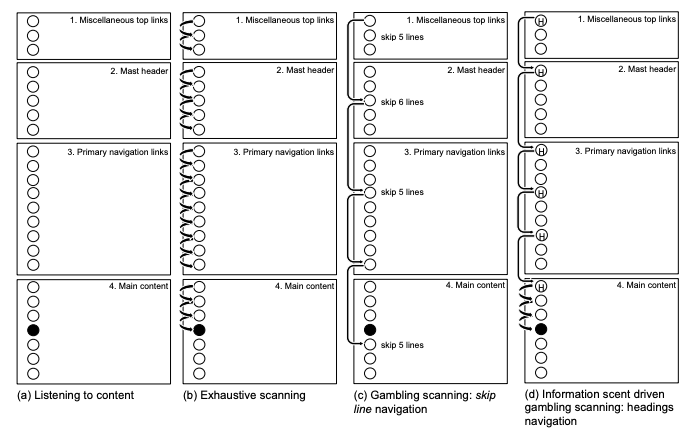}
\caption{Website exploration strategies employed by screen reader user as presented in \cite{vigo2013challenging}.}
\label{fig:screenreadernav}
\end{figure*}

Vigo and Harper's study \cite{vigo2013challenging} focused on tactics and coping mechanisms of vision impaired users when faced with Information Architecture barriers. Authors of \cite{vigo2013challenging} first, summarized the navigation techniques [Figure \ref{fig:screenreadernav}]  screen reader users employ when accessing a web page [Figure \ref{fig:steriolayout}] and then analysed the data of two independent empirical studies (one think aloud study and one observational study) with visually impaired users. Figure \ref{fig:screenreadernav}(a) showcases an effective yet inefficient way of navigation \cite{vigo2013challenging} were users listen to all the content without any interaction. Figure \ref{fig:screenreadernav}(d), however, describes an efficient way of navigation and a better alternative to gambling scanning [Figure \ref{fig:screenreadernav}(c)] when heading labels are suitable. In fact, an information scent driven gambling navigation [Figure \ref{fig:screenreadernav}(d)] switches from a gambling strategy to an exhaustive scanning when an information scent is deemed appropriate. 

 Vigo and Harper's \cite{vigo2013challenging} empirical findings challenge the information foraging scent driven navigation. In fact, their results summarized screen reader users coping mechanisms into nine different tactics and showed, for instance, that when faced with extreme situations (feeling lost) users deliberately choose to click on low scented links in an attempt to escape the problematic situation and ``backtrack to a shelter'' \cite{vigo2013challenging}.   

\paragraph{Online Communities (Collaborative) Websites} Authors of \cite{lachner2018culturally} and \cite{gregoriades2015combining} both investigated online communities through a design and evaluation process. For instance, Lachner et al. \cite{lachner2018culturally} combined the IA and Hofstede et al. \cite{hofstede2011dimensionalizing} power distance cultural dimension to first design two Q\&A websites (one targeted to the German community and another to the Vietnamese community) and then evaluated the two designs with both German and Vietnamese participants. Results of the think aloud were inline with their design hypothesis. 

\boldification{Educational Websites}
\paragraph{Educational websites} Articles related to educational websites focus on three difference niches: Online education tools, Open Educational Resources (OER) and academic websites. 

Even though \cite{buehler2016accessibility} and \cite{doubleday2013use} both investigated online education tools, their studies focused on the perspective of different target users.  
Buehler et al. \cite{buehler2016accessibility} provided a field study observation of the IA barriers faced by young adults with intellectual disability. Results of their study provided a list of recommendations for developers and researchers to help address the technological barriers. Authors of \cite{doubleday2013use} leveraged the input of current students in an iterative participatory design process to inform the design of a new online curriculum. The study employed both card sorting techniques and user studies to respectively design and evaluate the new curriculum in an iterative way. 

Leinonen et al. \cite{leinonen2010information} described a case study of an open educational resource website (LeMill). The paper focused on the educator's(teacher) perspective. They constructed the structure of LeMill's website using participatory design sessions and focused on an IA that encourages sharing and collaborating on learning materials. In contrast with LeMill's study, authors of \cite{navarrete2015evaluating} evaluated OER's adequacy for user with disabilities at the receiving end of the spectrum. In fact, Navarrete and Lujan-Mora \cite{navarrete2015evaluating}  analytically compared the findability on three OER websites using accessibility conditions, usability guidelines and IA principles. Their results advocate for a more consistent IA across OER platforms. 

The last study relevant to the educational milieu investigated the effect of the IA of an academic website \cite{gullikson1999impact}. In order to achieve their goal, Gullikson et al. \cite{gullikson1999impact} conducted a user study with 24 participants from six different universities. The academic website received a failing grade from the participants. Results showed that
the most frequently mentioned issues were related to the organization scheme (variable with the lowest user rating). While participants were satisfied with the content of the website, they reported confusing labels:
\begin{quote}
``kind of hard to know if something is under departments or academic'' \cite{gullikson1999impact} 
\end{quote}


\subsubsection{Computational Modeling, Automation and Mining}
Out of all the papers that fall under the website category, 27\% applied the IA lens for computational modeling, automation and mining. 
\paragraph{Computational Modeling}

\begin{figure}[hbtp]
  \centering
   \subfloat[A $4 X 2$ architecture with clear link labels]{{\includegraphics
   [width=.8\columnwidth]{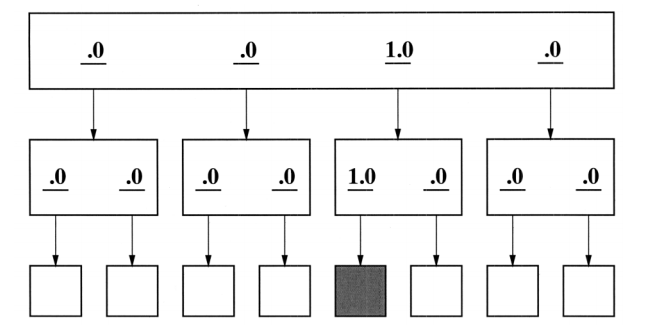} }}

    \subfloat[A $4 X 2$ architecture with some umbiguity added to the link labels.]{{\includegraphics[width=.8\columnwidth]{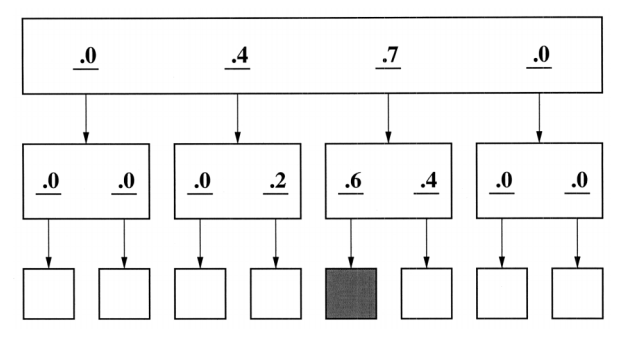} }}
      
   \subfloat[A $4 X 2$ architecture with misleading link labels.]{{\includegraphics[width=.8\columnwidth]{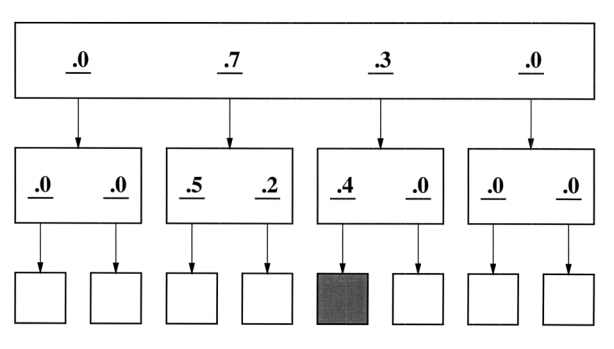} }}
   \caption{Three examples of different levels of label ambiguity in a $4 X 2$ architecture as presented in \cite{miller2004modeling}. The colored box corresponds to the target page.}
   \label{fig:LabelsLevels}
    \qquad
\end{figure}

Both \cite{miller2004modeling} and \cite{schaik2015automated} used computational modeling to investigate website Information Architecture and validated their analysis with user studies.
Even though \cite{miller2004modeling} and \cite{schaik2015automated} use two different computational models, there findings are consistence in that they both showed that an optimal navigation depends on the quality of the labels used even more so they the structure of a page \cite{miller2004modeling}. 

Authors of \cite{miller2004modeling} built the MESA model (Method for Evaluating Site Architecture) to simulate users navigation by varying both the site structure and labels ambiguity. Figure \ref{fig:LabelsLevels} displays three examples of a $4 X 2$ architecture with different levels of label ambiguity. The numbers [Figure \ref{fig:LabelsLevels}] correspond to the perceived relevance of link labels. For example, the number $1.0$ represents a high perceived relevance.

Miller and Remington's findings \cite{miller2004modeling} showed that there is an interaction between link labels and the structure of a site. In fact, a two-level structure (e.g $18 X 32$) resulted in a shorter navigation time compared to three-level structure (e.g $ 7 X 7 X 7 $) when all but bottom-level labels are ambiguous enough to require backtracking. In contrast, when the compared structure have clear labels the opposite is observed. Miller and Remington's findings \cite{miller2004modeling} emphasize on the importance of a navigation system with clear labels even more so then one with a perfect structure: 
\begin{quote}
    ``...  a top level with just a half dozen links could be part of an effective structure if the top level links reliably led the user to the next level.''\cite{miller2004modeling}
\end{quote}

In contrast, authors of \cite{schaik2015automated} extended the Comprehension-Based Linked Model of Deliberate Search plus(CoLiDeS+). CoLiDeS+ addresses some of the limitation the MESA model \cite{miller2004modeling} and the SNIF-ACT model by taking into consideration the context (e.g headers) (added in CoLiDeS model) and the navigation history (Specific to CoLiDeS+). The resulting model is an enhanced CoLiDeS+ \cite{schaik2015automated} that: 
\begin{itemize}
    \item Segments the current page and identifies two main sections (sections with headings and headingless sections that contain links).
    \item Considering the set of sections with heading: the model identifies the section whose heading displays the highest cosine (Latent Semantic Analysis (LSA) cosine) similarity with the goal and selects the link within that section that has the highest cosine similarity with the goal. 
    \item Considering the set of headingless sections, identifies the link that has the highest cosines value.
    \item From the two identified links the one having the highest cosine value is traversed.
\end{itemize}   
The cosine similarity of a link under a heading is equal to the ``heading cosine + the link cosine''. The cosine similarity would then be at least as high as it could be without the heading label. Moreover, for areas under a heading: The heading cosine similarity value is set as a cut off for all links under that heading. Thus, the ``heading cosine + the link cosine'' should always be greater or equal to the heading cosine.

A first goal driven simulation on two different websites (an academic subsite and a large intranet site), shed some light on potential IA barriers: 
\begin{itemize}
    \item Concise link labels that don't carry enough meaning.
    \item Poor link labeling (``e.g click here'').
    \item A large number of possible destinations (competing links and headers).
    \item Heading and link labels are not usually worded based on user's information goal. 
    \item Domain specific abbreviations that are not fully written.
\end{itemize}

A second simulation (using the same goals as the previous simulation) on the improved version of the same two websites (academic subsite and a large intranet site) resulted in a 34\% and a 38\% increase in the goal success rate and showed a positive effect of the IA changes on the navigation. 

Additionally, a user study on the original vs. the improved version of the two websites (academic subsite and a large intranet site) showed that the users of the improved versions of both websites (experimental group) outperformed the users of the original versions (control group). Findings showed that the differences between the experimental and control group were statistically significant on task completion, task correctness, time on task, and task performance (correctness/time).

\paragraph{Automation and Mining}
Wiki is a collaborative website that allows user to edit, add and review content. This flexibility encourages knowledge sharing and collaboration but provides no guarantees when it comes to the content structure \cite{bittar2010accessible}. In fact, the structure of a Wiki is usually the results of user generated links. Bittar et al. \cite{bittar2010accessible} proposed an automatic Wiki structure generation using name spaces and IA best practices in order to provide a more consistent navigation system. 

\begin{figure}[bthp]
\centering
\graphicspath{ {figures} }
         \includegraphics[width=.43\textwidth]{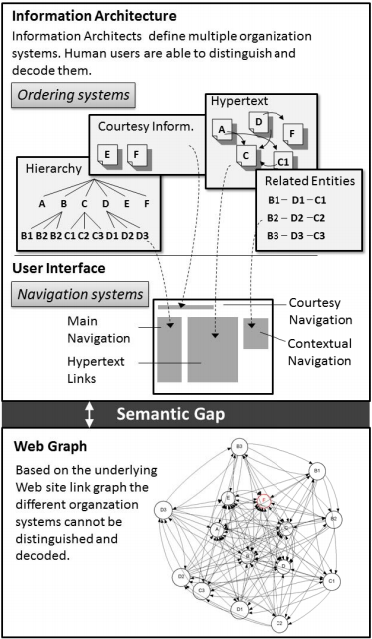}
\caption{The gap between the web graph and the way human perceive the Information Architecture as presented in \cite{keller2011beyond}.}
\label{fig:mininggap}
\end{figure}

Humans are able to decipher fine grained organization scheme and detect entry pages and website boundaries in  ways that machines can't \cite{keller2012menuminer}\cite{keller2011beyond}. 
Authors of \cite{keller2011beyond}, for instance, describe the semantic gap between web graphs\textemdash where a node represents a page and an edge represents a hyperlink [Figure \ref{fig:mininggap}] and the more fine grained human understanding of IA. In fact, as depicted in Figure \ref{fig:mininggap} a lot of details (e.g, type of navigation and ordering system) are hidden in the coarse grained web graph model. In order to bridge the aforementioned gap, Keller and Nussbaumer \cite{keller2011beyond} analyzed the Navigation Structure Graph (NSG) by implementing a finer grained navigation element detection and employing continuity analysis to distinguish between the different types of navigation systems present. Results of their  prototype evaluation displayed a 78\% error free navigation element detection and a 44\% error free hierarchy detection.
While both \cite{keller2012menuminer} and \cite{keller2011beyond} structure extraction methods focus on static website, authors of \cite{zhao2017navigation} introduce a method that captures both the static and dynamic structure of a website. 

\subsection{The Not So Prevalent Categories}
All the paper in this section fall under the design and evaluation subcategory. 
\paragraph{Mobile Applications} IA in mobile applications represents 12\% of our final set of papers making it the second most prevalent category. Similarly to papers that fell under the website category, the mobile category includes research that investigate both (visual) IA and aural IA. Both Gatsou et al. \cite{gatsou2012novice} and Punchoojit and Hongwariton \cite{punchoojit2015children} conducted user studies to investigate how specific groups of users structure information using the open card sorting technique. On one hand, Gatsou et al. \cite{gatsou2012novice} focused on how novice users construct the Information Architecture of a mobile tablet application resulting in more uni-level groupings compared to the intended application structure. 
On the other hand, Punchoojit and Hongwariton \cite{punchoojit2015children} studied the similarities and differences in information structuring between adults and children when faced with
both concrete and abstract categories. 
Gross et al. \cite{gross2018exploring} designed and evaluated a non screen-centric aural navigation paradigm for vision impaired users. This study introduced a binary dichotomic aural information navigation scheme that leverages topical, usage based and user defined organizational scheme. Findings from their empirical evaluation reported an above average perceived usability (System Usability Scale (SUS) questionnaire) \cite{brooke1996sus} and a bellow average computed lostness \cite{albert2013measuring} for four out of the five administrated tasks. Similarly to Rohani et al. paper \cite{rohani2012back} in the website category, Yang et al. \cite{yang2012aural} introduced and investigated a topic and list based back navigation in a mobile application case.

\paragraph{IoT (Smart Toys)}
Salgado et al. \cite{de2019smart} bridges the gap of IA for usable security tools. In order to do so, they compared the structure scheme of a card sorting experiment with the nutrition label model \cite{kelley2009nutrition} commonly used in usable security. The final result is an IA inspired  prototype. 

\paragraph{Semantic Web}
The three papers in this category \cite{garcia2011publishing} \cite{brunetti2012improved} \cite{brunetti2013design} correspond to different iterations and improvement on an IA driven semantic web structuring application (Rhizomer). The goal of the aforementioned research is to tackle the challenges of structuring and organizing semantic data by leveraging existent IA components.
For instance, authors of \cite{brunetti2013design} adapted both the global and local navigation menu by dynamically taking into consideration the frequency of each menu class. Classes that are heavily instantiated have a more prominent positioning in the menu bar. Brunetti \cite{brunetti2013design} also used site maps (based on the original hierarchy of the data), site indexes (alphabetical order) and tree maps as an alternative navigation components. The study's empirical findings showed that, navigational menus overcame the use of other components for finding highly instantiated items. The implemented summarized and complete sitemaps, however, had some limitation associated to them. In fact, they required the users to understand the data structure in order to use them efficiently. 

\paragraph{Ubiquitous Technology}
Oleksik et al. \cite{oleksik2013towards} ran an exploratory co-design study with both faculty and PhD student participants to investigate the accessibility, use and reuse of information across devices in a research environment. The study findings emphasized the need for expending the traditional Information Architecture of file systems for ubiquitous technology and promoting a more fine grained, multi-format information representation in a way that makes referencing and linking between information units independent of their format and their level of granularity (file vs sub-file).   

\paragraph{Information Visualization}
Li et al. \cite{li2017comparing} studied the connection between Information Architecture and information visualization. The study \cite{li2017comparing} leveraged IA principles to compares three different information visualization (a published visualization and two experimental ones) targeted toward high school students. The compared visualization techniques use three different Information Architecture models (tunnel, matrix and hybrid (Matrix with hierarchy)). The authors used the eight IA principles introduced by Dan Brown \cite{brown2010eight} to investigate the potentials and pitfalls of the aforementioned techniques. The results show that a hybrid method that combines both the rich connection of a matrix IA with a hierarchical structure enables a drill down from groups to individual pieces of information in a way that is simpler for users to understand. In a broader sense, their findings argue that IA principles could be used to improve information visualization tools. 

\paragraph{Usability Questionnaire}
Yang et al. \cite{yang2012deep} proposed a usability questionnaire using IA to capture actionable users feedback. The main goal of this questionnaire is to go beyond the detection of usability issues by distinguishing the underlying design issues. DEEP\textemdash the designed questionnaire\textemdash is composed of 19 questions spanning five dimensions: content, Information Architecture, navigation, layout, and visual guidance. 
\section{Discussion}
\label{sec:discussion}
The purpose of this Systematic Literature Review is to provide researchers with an overview of the current state of the art of IAinHCI in an attempt to uncover research opportunities that can leverage the vast IA knowledge base. 

In the last decade, Information Architecture was primarily tailored toward websites and more specifically toward the design and evaluation of websites. These results are consistent with the seminal work of Morville and Rosenfeld \cite{morville2006information} and Wodtke and Govella \cite{wodtke2011information}. In fact, Rosenfeld and Morville's work  \cite{rosenfeldinformation} was targeted toward ``applying the principles of architecture and library science to web site design'' This statement has long been taken as a given, narrowing the application domain to web environments \cite{dillon2005information}. In recent years, however, an emergence of less prevalent categories (e.g, information visualization) shows promising results as to the potential of applying IA to domains other then the web [Figure \ref{fig:categperyear}]. 

Even though, some research investigated the Information Architecture of collaborative online communities (e.g, Q\&A websites) [Figure \ref{fig:D&Ewebsite}], there appears to be little to no research that leverages IA in Open Source platforms. Due to its challenging onboarding process, the Open Source domain has been receiving a lot of attention lately  \cite{steinmacher2016overcoming}\cite{steinmacher2015social}\cite{balali2018newcomers}\cite{mendez2018open}. This raises the question of whether the way information is structured, organised and labelled has an impact on contributors experience in general and newcomers onboarding in particular. Mining and Computational Modeling could help automatically detect problematic IA elements in Open Source environments, building the first stepping stones toward an IA metric.  

Moreover, in the absence of an agreement on a clearly established IA definition, bounding work exclusively to IA remains a challenge especially knowing that information structure and navigation have been around for a long time in different disciplines \cite{dillon2005information}. In addition, the absence of validated IA principles makes the delimitation of the impact of Information Architecture harder to grasp, yet opens an opportunity of defining clear IA principles and guidelines in the same way typography and color theory provide principles for graphic design \cite{brown2010eight}. This could be particularly useful for User Interface (UI) practitioner and information architects.

\section{Threats to validity}
\label{sec:threat}

Zhou et al. \cite{zhou2016map} map the threats to validity mentioned in Software Engineering Systematic Literature Reviews. The four most frequently encountered threats to validity are: 
\begin{itemize}
    \item Bias in study selection. 
    \item Bias in data extraction. 
    \item Inappropriate or incomplete search
terms in automatic search. 
    \item Non comprehensive venues or databases. 
\end{itemize}

To avoid selection bias we followed the SLR process as defined by \cite{kitchenham2004procedures}, used a predefined list of inclusion/ exclusion criteria and documented the reasons for any paper removal at every step. Whenever a paper's abstract was unclear or had missing information, we kept the paper for further scanning during the full text read. In addition to that, we complemented the SLR process with a backward reference search \cite{skoglund2009reference} in order to capture previous relevant publications. 

To help insure that the data extraction was unbiased, we established our central research question and the list of data to be extracted from the paper prior to beginning the SLR process. All the papers and extracted data were stored in an excel sheet. 

The inappropriate or incomplete search terms in automatic search was mitigated by (1) running pilot searches to decide on a search string and its scope and (2) manually screening for papers that are relevant to using IA in the HCI field. The non comprehensive venues or database threat was mitigated by avoiding the selection of individual conferences and rather querying both the ACM and IEEE databases which were relevant to the domain of our research question.  

\section{Conclusion}
\label{sec:conclusion}
In this paper, we filtered and assessed 311 papers using a SLR that spanned a decade of research on Information Architecture in the context of HCI. We then, complemented the SLR process with a Background Reference Search. The two aforementioned methods resulted in a total of 33 papers that covered seven different categories.

Our results were consistent with seminal work in Information Architecture. Over the past decade, Information Architecture in the domain of HCI was most commonly applied for the design and evaluation of web environments with 67\% of IAinHCI papers falling under the website category. However, our findings also showed that different flavors of Information Architecture ((visual)IA and aural IA) were used in different topics (IoT, ubiquitous technology, websites, information visualization...) and for a diverse set of users. This opens the opportunity for additional theoretical and technical research in multiple disciplines.

With the growing amount of information rich environment, Information Architecture seems assured of a prosperous future. Its multidisciplinary roots provide a rich knowledge base that has yet to unravel its full potential providing multiples opportunities for this domain to expend and grow.

\bibliographystyle{IEEEtran}
\bibliography{main} 
\balance
\end{document}